\documentclass[journal]{IEEEtran}
\usepackage{epsfig,times,amsfonts,amsmath,amssymb}
\usepackage{amsmath,amssymb}
\usepackage{cite}
\usepackage{graphicx}
\usepackage{caption2}
\usepackage{verbatim}
\begin{document}

\markboth{To appear in IEEE Transactions on Communications}%
{Shell \MakeLowercase{\textit{et al.}}: Soft-Decision-Driven Channel
Estimation for Pipelined Turbo Receivers}

\title{ \huge Soft-Decision-Driven Channel Estimation \\ for Pipelined Turbo Receivers}
\author{Daejung~Yoon,~\IEEEmembership{Student Member, IEEE}
        and~Jaekyun~Moon,~\IEEEmembership{Fellow, IEEE}
\thanks{This work was supported in part by the NSF under Theoretical
Foundation grant 0728676. Daejung Yoon is with Dept. of Electrical
and Computer Engineering, University of Minnesota at Minneapolis, MN
55455, USA (email:yoonx053@umn.edu), and Jaekyun Moon is with Dept.
of Electrical Engineering, Korea Advanced Institute of Science and
Technology, Daejeon, Republic of Korea (email: jmoon@kaist.edu)}}
\maketitle

\begin{abstract}
\noindent {\it We consider channel estimation specific to turbo
equalization for multiple-input multiple-output (MIMO) wireless
communication. We develop a soft-decision-driven sequential
algorithm geared to the pipelined turbo equalizer architecture
operating on orthogonal frequency division multiplexing (OFDM)
symbols. One interesting feature of the pipelined turbo equalizer is
that multiple soft-decisions become available at various processing
stages. A tricky issue is that these multiple decisions from
different pipeline stages have varying levels of reliability. This
paper establishes an effective strategy for the channel estimator to
track the target channel, while dealing with observation sets with
different qualities. The resulting algorithm is basically a linear
sequential estimation algorithm and, as such, is Kalman-based in
nature. The main difference here, however, is that the proposed
algorithm employs puncturing on observation samples
to effectively deal with the inherent correlation
among the multiple demapper/decoder module outputs that cannot easily be removed by
the traditional innovations approach. The proposed algorithm
continuously monitors the quality of the feedback decisions and
incorporates it in the channel estimation process. The proposed
channel estimation scheme shows clear
performance advantages relative to existing channel estimation techniques.}
\end{abstract}

\section{Introduction}\label{Intro}

Combining the multiple-input multiple-output (MIMO) antenna method
with orthogonal frequency division multiplexing (OFDM) and spatial
multiplexing is a well-established wireless communication technique.
Bit-interleaved coded modulation (BICM) \cite{Caire98} used in
conjunction with MIMO-OFDM and spatial multiplexing (SM) is
particularly effective in exploring both spatial diversity and
frequency selectivity without significant design efforts on
specialized codes \cite{Tonello00, Park03}. Turbo equalization \cite{TurboEQ}, also
known as iterative detection and decoding (IDD) in wireless
applications \cite{Tuchler02_1}, is well-suited for BICM-MIMO-OFDM
for high data rate transmission with impressive performance
potentials \cite{Koetter04, Tuchler02_1}.

A critical issue in realizing the full performance potential of a MIMO-OFDM
system is significant performance degradation due to
imperfect channel state information (CSI). The detrimental impact of
imperfect CSI on MIMO detection is well known (see, for example,
\cite{Huang03, Huang13}) and continues to be a great challenge in
wireless communication system design. Previous works have identified desirable
training patterns or pilot tones for estimating channel responses
for MIMO systems \cite{Li99, Li98, Li02, Ma04, Hassibi00}. However, with these methods
the achievable data rate is inevitably reduced, especially when the number of
channel parameters to be estimated increases (e.g., caused by an
increased number of antennas).

Decision-directed (DD) channel estimation algorithms can be applied
to the turbo receivers to improve channel estimation accuracy
\cite{Sandell98, Gao05, Deng03, Loncar02}. However, inaccurate
feedback decisions degrade the estimator performance
\cite{Tuchler02}. Maximum-a-posteriori (MAP)-based DD algorithms
discussed in \cite{Deng03, Gao05} can improve the estimation
accuracy, but they require additional information like the channel
probability density function. The DD channel estimation algorithm
jointly working with IDD has been actively researched
\cite{Kobayashi01,Wautelet07,Khalighi06,Song04,Song02}. Among the
existing research works, several papers have been devoted to
iterative expectation-maximization (EM) channel estimation
algorithms using extrinsic or a posteriori information fed back from
the outer decoder \cite{Kobayashi01,Wautelet07,Khalighi06}. Although
the traditional EM-based estimation algorithms typically show outstanding
performance, the heavy computation complexity and the iteration
latency can be problematic for many practical applications. While
an approximation scheme as discussed in \cite{Kobayashi01} can
reduce complexity, the
performances of these approaches suffer from performance degradation
as the number of antennas increases
\cite{Wautelet07}. Also, the EM estimation algorithms need to be
aided by pilot-based EM algorithms to guarantee a good
performance\cite{Kobayashi01,Khalighi06}.

As an alternative approach to iterative EM channel estimation,
Kalman-based channel estimators have been discussed that are
effective against error propagation \cite{Song04,Song02}. The
authors of \cite{Song04,Song02} have introduced a soft-input channel
estimator that adaptively updates the channel estimates depending on
feedback decision quality. The soft-input channel estimator of
\cite{Song04} evaluates the feedback decision quality by tracking
the noise variance that includes the potential soft-decision error
impact in its effort to improve the update process for the Kalman
filter.

In the present work, we develop a Kalman-based channel estimator for
MIMO-OFDM based on a specific pipelined turbo equalizer receiver
architecture. Before setting up the Kalman estimator, a novel method
for reducing decision error correlation is introduced. The proposed
method constructs a refined innovation sequence by irregularly
puncturing certain soft decisions that are deemed to be correlated
with the previous decisions. The resulting algorithm is basically a
linear sequential estimation algorithm and, as such, is Kalman-based
in nature. We also weigh the estimated channel responses in the
detection process according to the quality of the estimation.

A critical issue in turbo receiver design is long processing latency
due to inherent iterative processing of information. Pipelined
architecture reduces the latency and improves processing throughput
in turbo receivers and thus is the prevailing choice of the
implementation architecture \cite{Abbasfar, Lee03}. One interesting
feature of the pipelined turbo equalizer is that multiple sets of
soft-decisions become available at various processing stages. A
tricky issue is that these multiple decisions from
different pipeline stages have varying levels of reliability.
Therefore, an adequate optimization strategy is required for the
estimator to track the target channel while dealing with observation
sets with different qualities. An optimum channel estimator is
derived based on this principle for the pipelined turbo receiver.

In demonstrating the viability of the proposed schemes, a
SM-MIMO-OFDM system is constructed to comply with the IEEE 802.11n
high speed WLAN standard \cite{IEEE11n}. Section \ref{MIMOOFDM}
discusses the channel and system model, and briefly touches upon the
high-throughput pipelined IDD architecture. Section \ref{CE}
discusses the method to set up an improved innovation sequence via
puncturing. Next, the proposed soft-DD Kalman-based channel
estimation methods are presented in section \ref{CE}. Mean squared
error (MSE) analysis is provided in Section \ref{MSE} that validates
the performance merits of the proposed schemes. In Section \ref{PE},
the convergence behavior is investigated via the extrinsic
information transfer (EXIT) charts \cite{tenBrink00}, and packet
error rate (PER) simulation results are presented for performance
evaluation. Finally conclusions are drawn in Section \ref{con}.

\section{Channel and System Model}\label{MIMOOFDM}

We assume a SM-MIMO-OFDM transmitter where a data bit sequence is
encoded by a convolutional channel encoder, and the encoded bit
stream is divided into $N_t$ spatial streams by a serial-to-parallel
demultiplexer. Each spatial
stream is interleaved separately, and the interleaved streams are
modulated using an $M$-ary quadrature amplitude modulation ($M$-QAM)
symbol set $\mathcal {A}$ based on the Gray mapping. Since $Q$ binary bits form an
$M$-QAM symbol, a binary vector ${\bf{b}} = [b_0 ,b_1 , \cdots
,b_{QN_t - 1} ]^T$ is mapped to a transmitted symbol vector
$\mathbf{s}=\left[s_1,s_2,\cdots,s_{N_t}\right]^T$ (with $s_i \in \mathcal
{A}$), taken from the set $\mathcal {A}^{Nt}$, a Cartesian product of $M$-QAM constellations. The
$M$-QAM symbol sequence in each spatial stream is transmitted by an
OFDM transmitter utilizing a fixed number of frequency subcarriers.
For a particular subcarrier for the $n^{th}$ OFDM symbol, the
received signal at the discrete Fourier transform (DFT) output can
be written as
\begin{equation}
\mathbf{z}_n=\mathbf{H}\mathbf{s}_n+\mathbf{n}_n \ , \label{eq:z}
\end{equation}
where $\mathbf{z}_n=\left[z_1(n),z_2(n),\cdots,z_{N_r}(n)\right]^T$
is the received signal vector observed at the $N_r$ receive
antennas, and $\mathbf{H}$ is the channel response matrix associated
with all wireless links connecting $N_t$ transmit antennas with
$N_r$ receive antennas antennas, and $\mathbf{n}_n$ is a vector of
uncorrelated, zero-mean additive white Gaussian noise (AWGN) samples
of equal variance set to ${\mathcal N}_o$.

\begin{figure}
\centering
\includegraphics[width=3.5in]{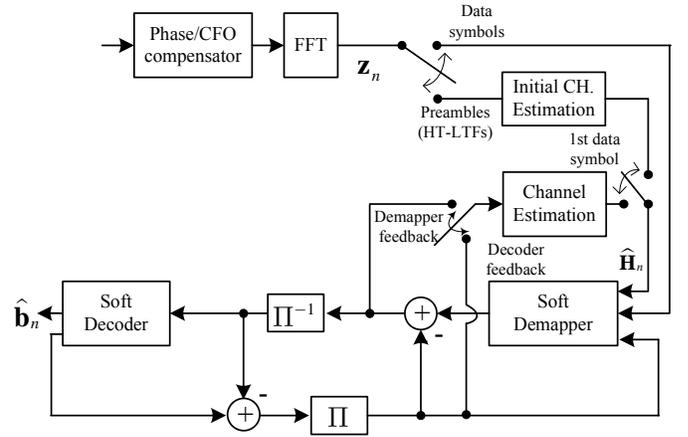}
\centering\captionstyle{flushleft} \caption{Block diagram of the
turbo receiver and the soft-decision-directed channel estimator}
\label{fig:IDD}
\end{figure}

\begin{figure*}
\centering
\includegraphics[width=4.9in]{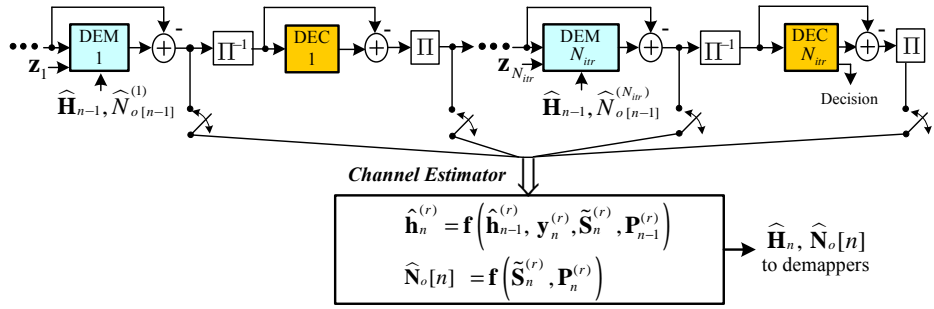}
\caption{Block diagram of the proposed optimum channel estimation
algorithm geared to the pipelined IDD} \label{fig:pipeline_opt}
\end{figure*}

The IDD technique of \cite{Koetter04} that performs turbo
equalization for MIMO systems is assumed at the receiver. The
extrinsic information on the coded-bit stream is exchanged in
the form of log-likelihood ratio (LLR) between the soft-input
soft-output (SISO) decoder and the SISO demapper as shown in
Fig.~\ref{fig:IDD}. The demapper takes advantage of the reliable
soft-symbol information made available by the outer SISO decoder. A
soft-output Viterbi algorithm (SOVA) is used for the SISO decoder
implementation \cite{Hagenauer89}. Each data packet transmitted
typically contains many OFDM symbols, and they are processed
sequentially by the demapper and the decoder as they arrive at the
receiver. The feedback decisions used for
channel estimation must be interleaved coded-bit decisions. The
extrinsic information from the demapper are rearranged accordingly
and made available to the channel estimation block.
\begin{figure}
\centering
\includegraphics[width=3.5in]{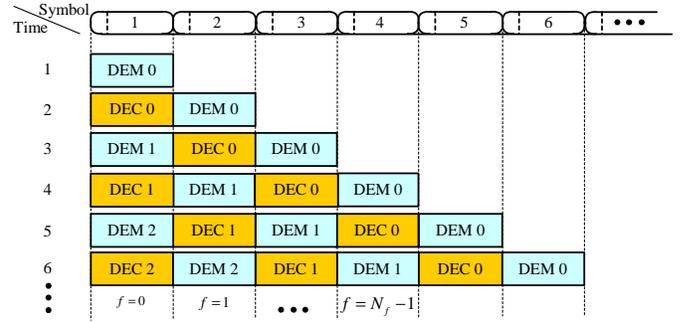}
\caption{OFDM-symbol processing procedure in the pipelined IDD}
\label{fig:timing}
\end{figure}
The pipelined architecture is adopted to reduce the iteration
latency ~\cite{Abbasfar, Lee03}. Fig.~\ref{fig:timing} illustrates
OFDM symbols processed in the pipelined IDD, and
Fig.~\ref{fig:pipeline_opt} shows the structure of the pipelined IDD
receiver and its interface with the channel estimator. Multiple
demapper-decoder pairs process multiple OFDM symbols at different
iteration stages. Let $N_{itr}$ denote the number of the IDD
iterations required to achieve satisfactory error rate performance.
The $N_{itr}$-stage pipelined IDD receiver is equipped with
$N_{itr}$ demappers and $N_{itr}$ decoders that are serially
connected as in Fig.~\ref{fig:pipeline_opt}. The decoder forwards
its extrinsic information output to the demapper in the next
iteration stage. Simultaneously, the demapper and the decoder in the
previous iteration stage start to process a new OFDM symbol. The
pipelined IDD operation is functionally equivalent to the original
IDD scheme~\cite{Abbasfar}.
The extrinsic LLRs released from the pipelined demappers and
decoders are utilized for the channel estimation. Let $N_{sym}$
denote the number of total OFDM symbols in a packet and $N_f$ the
number of the feedback symbols available for channel estimation.
Note, however, that not all $N_f$ symbols are used for the
estimation. If the receiver requires $N_{itr}$ IDD iterations, then
a maximum of $2N_{itr}$ OFDM symbols are processed in the pipelined
IDD receiver as illustrated in Fig.~\ref{fig:pipeline_opt}. Because
the LLR outputs from the initial demapper and decoder have low
reliability, they are not used for the channel estimation. Let index
$n$ indicate the time. In this pipelined IDD setup, when $2\le
n\le2N_{itr}$, the channel estimator can get ($n-2$) feedback
decisions (i.e. $N_f=n-2$). When the number of the processed symbols
increases to $2N_{itr}$ ($2N_{itr} \le n \le N_{sym}$), $N_f$ is
equal to $2 N_{itr}-2$. After all the OFDM symbols in the packet
have arrived at the receiver front-end, it will take sometime until
all symbols will clear out of the pipeline. For $n \ge N_{sym}$,
$N_f$ is equal to $N_{sym} + 2N_{itr} - n$.

\section{Sequential and Soft-Decision-Directed Channel Estimation}\label{CE}

\subsection{Derivation of the Kalman-Based Sequential Channel Estimation Algorithm}\label{Kalman}

The sequential form of the estimator is useful in improving the
quality of the channel estimate as the observed symbols arrive in a
sequential fashion, as OFDM symbols do in the system of our
interest. It is assumed that the channel is quasi-static over $N_f$
OFDM symbol periods. For the pipelined IDD receiver at hand, the
observation equation is set up at the $r^{th}$ receiver (RX) antenna
as
\begin{equation}
\mathbf{z}^{(r)}_n=\mathbf{S}_n \mathbf{h}^{(r)}+\mathbf{n}^{(r)}_n,
\ \label{eq:Mz}
\end{equation}
where $\mathbf{z}_n^{(r)}$ is the ${N_f \times 1}$ received signal
vector, $\mathbf{S}_n$ is a ${N_f \times N_t}$ matrix,
$\mathbf{h}^{(r)}$ is a ${N_t \times 1}$ vector that is a
multi-input-single-output (MISO) channel vector specific to the
${r^{th}}$ RX antenna. The goal is to do a sequential estimation of
$\mathbf{h}^{(r)}$ as $n$ progresses. The estimation process is done
in parallel to obtain channel estimates for all ${N_{r}}$ RX
antennas. With an understanding that we focus on a specific RX
antenna, the RX antenna index $r$ is dropped to reduce notation
cluttering.

A mean symbol decision $\tilde s$ is defined as the average of the
constellation symbols according $\tilde s = \tiny{
\sum\nolimits_{\tiny{s_i \in {\mathcal {A}}}} {s_i P(s_i )}}$, where
$P(s_i)$ is the ``extrinsic probability" obtained from a direct
conversion of the available extrinsic LLR.

\subsubsection{Innovation Sequence Setup}\label{innovation}

The pipeline architecture can be viewed as a buffer large enough to
accommodate $N_f$ OFDM symbols, but we take into account in our
channel estimator design the different levels of reliability for the
soft decisions coming out of the demapper or decoder modules at
different iteration stages. First, defining the soft decision error
$\mathbf {E} \buildrel \Delta \over =\mathbf{S}-\widetilde{\mathbf
S}$, (\ref{eq:Mz}) can be rewritten as
\begin{equation}
\mathbf{z}_{n}=\{\widetilde{\mathbf S}_n + \mathbf{E}_n\}
{\bf{h}}+\mathbf{n}_n \label{eq:ZE}.
\end{equation}
Note that this type of  soft decision representation has been used
previously ~\cite{Song04}. We emphasize, however, that unlike in
\cite{Song04}, our derivation of a linear sequential estimator is
based on the attempt to explicitly generate the innovation
sequence. As will be clear in the sequel, this approach has led us
to a realization that the standard steps taken to generate the
innovations do not work in our set up; this in turn allowed us to devise
corrective measures. Let us first see if we can find
$\mathbf{x}_{n}$, the innovations of $\mathbf{z}_{n}$ (i.e., the
whitened sequence that is a causal, as well as a casually
invertible, linear transformation of $\mathbf{z}_{n}$). We write:
\begin{eqnarray}
\mathbf{x}_{n}    &\buildrel \Delta \over =&  \mathbf{z}_n-
\widetilde{\mathbf S}_n \hat {\mathbf{h}}_{n-1} \label{eq:yr} \\
&=& \widetilde{\bf{S}}_n ({\bf{h}} - \hat{\bf{h}}_{n - 1} ) +
{\bf{E}}_n {\bf{h}} + {\bf{n}}_n \label{eq:y}.
\end{eqnarray}
Ideally, the vector sequence $\mathbf{x}_{n}$ would represent an
innovation sequence in the sense that any given component of the
vector ${\bf{x}}_{n-k}$ is orthogonal to any component of
${\bf{x}}_n$ as long as $k\ne0$. In this scenario we would have
\begin{eqnarray}
E\left[ {x_{n - k} [i]x_n^* [j]} \right] = E\left[ {{\bf{e}}_{n - k}
[i]\widehat{\bf{h}}_{n - k - 1} {\bf{h}}^H {\bf{e}}_n^H [j]} \right]
+  \mathcal{N}_o \nonumber  \\
=\left\{ {\begin{array}{*{20}c}{\sum\nolimits_{t = 1}^{N_t }
{\hat\rho_{n - 1}^{(t)} } \sigma _s^2 [t,i]}
+  \mathcal{N}_o & {{\rm{when}}\,\,k = 0\,\,{\rm{and}}\,\,i = j}  \\
\epsilon (\approx 0)& {{\rm{otherwise}}} , \\
\end{array}} \right.  \label{eq:Eyy}
\end{eqnarray}
where ${\bf{e}}_n[i]$ indicates the $i^{th}$ row vector
${\bf{E}}_n$, ${\hat\rho_{n - 1}^{(t)} } \buildrel \Delta \over
= E[\hat h_{n - 1}^{(t)} h^{*(t)} ]$ and $\sigma_s^2\buildrel \Delta
\over =E[|s - \tilde s|^2]$, the symbol decision error
variance. The superscript `$H$' and the symbol `$*$' denote the Hermitian
transpose and the complex-conjugate,
respectively. In deriving (\ref{eq:Eyy}), we assumed: $ E[{x}_{n -
k}[i] ({\bf{h}} - \widehat{\bf{h}}_{n - 1} )^H ] = \bf0$,
$E[s{[i]}\,e^*{[j]} ] = 0$ for any $k$, $i$ and $j$. In order for
this to be true, though, the following must hold:
\begin{enumerate}
\item Links in the MISO channel are uncorrelated.
\item The channel estimate and decision error are independent.
\item The decision errors are uncorrelated. \\
(i.e. $E[e_{n-k}[i] e_n^*[j]]=\epsilon, k \ne 0$ or $i \ne j$)
\end{enumerate}
Under these three assumptions, the vector ${\bf {x}}_n$ reasonably
represents an innovation sequence.

\subsubsection{Innovation Sequence with Punctured Feedback}\label{innovation}

Assumption (1) is reasonable, if the RX antennas maintain reasonable physical separation.
However, assumptions (2) and (3)
are not convincing. A poor channel estimate generates a poor decision,
which in turn affects the ability to make reliable channel estimation.
This makes both assumptions (2) and (3) invalid.
As such, the Kalman filter is not optimum any more, and the
correlated error circulates in the IDD and channel estimator loop. Our
goal here is to provide a refined innovation sequence to reduce this
error propagation.
\begin{figure}
\centering
\includegraphics[width=3.5in]{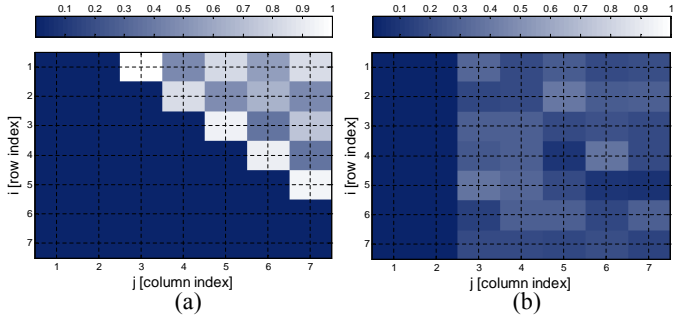}
\caption{Correlations in the `innovation' sequence:  (a)
$E[{\bf{x}}_{n - 2} {\bf{x}}_n^H]$ (b) $E[{\bf{y}}_{n - 2}
{\bf{y}}_n^{H}]$, ${c=0.8}$ (normalized by $E[|{y}_{n}[0]|^2]$,
averaging 50 erroneous packets)}\label{fig:corr}
\end{figure}
First, we observe that there is no significant correlation between
the decision errors of the demapper and decoder thanks to the
interleaver/deinterleaver. An issue is the demapper-demapper or
decoder-decoder output correlations for a given received signal
(OFDM symbol), especially when a packet is bad (i.e., certain tones
cause errors despite persistent IDD efforts). In the pipelined IDD
setup, it takes $n=2$ time steps for a demapper decision to shift to
the next-stage demapper, and likewise for the decoder outputs.
Consequently, components in observation vectors with even time
difference has correlation, as seen in Fig.\ref{fig:corr} (a)
between ${\bf x}_n$ and ${\bf x}_{n-2}$. In addition, we cannot
assume that the noise is random as long as identical observations
are reused during the iterative channel estimation.

This correlation in ${\bf{x}}_n$ is definitely problematic for any
Kalman estimator design. Imagine removing correlation in
${\bf{x}}_n$ using the Gram-Schmidt procedure:
\begin{eqnarray}
x_n^{'}[f]  = x_n[f]  - \frac{{\left\langle {x_n[f] ,\,\,\,x_{n -
2}[f-2] \,} \right\rangle }}{{\left| {x_{n - 2}[f-2] } \right|^2
}}x_{n - 2}[f-2] \label{eq:GS}
\end{eqnarray}
where $<a,b>$ denotes the inner product:
$<a,b>=Re(a)Re(b)+Im(a)Im(b)$. Now, (\ref{eq:Eyy}) can be rewritten
(for $k=2$ and dropping indices to simplify notation) as
\begin{eqnarray}
E\left[ {x_{n - 2} \,x_n^* } \right]\, &=& E\left[ {x_{n - 2}
\,x_n^{'*} } \right] + E\left[ {\left| {x_{n - 2}} \right|^2
\frac{{\left\langle {x_n,x_{n - 2} }
\right\rangle }}{{\left| {x_{n - 2} } \right|^2 }}} \right] \nonumber \\
&=& E\left[ {\left\langle {x_n ,\,x_{n - 2} } \right\rangle} \right]
\label{eq:inner}
\end{eqnarray}
which suggests that using only those samples of $x_n$ for which
${\left\langle {x_n ,\,\,\,x_{n - 2} \,} \right\rangle }\le \epsilon
$, where $\epsilon$ is an adjustable threshold, we can limit the
amount of correlation in the overall observation samples utilized.

Before delving into the proposed ``puncturing" process, we note that
 the amount of puncturing needs be decided judiciously,
 as removing observation samples also tends to ``harden" the decisions,
 making the overall system approach one of hard decision feedback, a situation we need to avoid.
Also, one may be tempted to use a more conceptually straightforward
approach of subtracting out the correlated component as suggested by
(\ref{eq:GS}) or its generalized version including subtraction of
less correlated components, but we had no meaningful success in reducing
correlated errors with approaches along this direction.

Equation (\ref{eq:inner}) suggests the following as a measure of correlation between the previous
demapper and current demapper outputs (or between the previous
decoder and current decoder outputs):
\begin{eqnarray}
\beta_n(f)  \buildrel \Delta \over = \left\langle {x_{n - 2}[f -
2],\,\,\,x_n[f] \,} \right\rangle. \label{eq:beta}
\end{eqnarray}
Now redefine $N_d$ as the number of components among ${x}_n(f)$'s
satisfying a threshold condition of
\begin{eqnarray}
| \beta _n (f)|\,\, \le \,\,\,c\,\mathcal{N}_o. \label{eq:threshold}
\end{eqnarray}
With this condition, let index $d$ now denote the number of selected
components among $N_f$ feedback symbols (i.e. $d=0,...,N_d-1$). The
constant $c$ ($\ge0$) is an important parameter that controls the
puncturing threshold. An improved innovation sequence ${\bf{y}}_n$
can be written as
\begin{eqnarray}
{\bf{y}}_n  = {\bf{G}}_n{\bf{x}}_n , \label{eq:reinnovate}
\end{eqnarray}
where ${\bf{G}}_n$ is defined as a $N_d \times N_f $ puncturing
matrix. For the $d^{th}$ row vector ${\bf g}_n^{(d)}$ of the
puncturing matrix, elements are given as
\begin{equation}
{ g}_n^{(d)} [f] = \left\{ {\begin{array}{*{20}c}
   {1,} & {if\,\beta _n (f)\le c \mathcal{N}_o\, or\, d=f=0 \,or\, d=f =1}\\
   {0,} & {{\rm{otherwise}}}.  \\
\end{array}} \right. \label{eq:puncmat}
\end{equation}
Note $x_n[0]$ and $x_n[1]$ are new input elements from the first
demapper and decoder outputs, which are automatically included in
the refined innovation vector. As long as the observations are
reused during the iterative process, the noise correlation is also
problematic in the channel estimation. To resolve this issue, a
scaled noise variance is adopted as a threshold criterion to judge
minimum correlation, because the noise variance term in
(\ref{eq:Eyy}) is inevitable. Highly correlated signal and noise
components are punctured out depending on the constant $c$.

It is insightful to consider a simple argument based on random
puncturing. Suppose the observation samples are dropped in a random
fashion. Then, the element $x_{n - 2}[f-2]$ can of course be
excluded from ${\bf{y}}_{n-2}$ by puncturing, and so can $x_n[f]$
from ${\bf{y}}_{n}$. With random puncturing, the innovation process
on each element can be analyzed as (dropping index $f$)
\begin{eqnarray}
\begin{array}{l}
E\left[ {y_{n - 2} y_n^{*}} \,\, \right] = E\left[ {g_{n - 2} x_{n - 2} x_n^* g_n } \right]   \\
\,\,\,\,\,= P\left( {g_{n - 2}  = 1} \right)E\left[ {x_{n - 2} x_n^* } \right]P\left( {g_n  = 1} \right) \\
\displaystyle \,\,\,\,\,=\frac{{N_{d}^{(n - 2)} }{N_{d}^{(n)}
}E\left[ {x_{n - 2} x_n^* }\right]}{\left({N_{f}^{(n - 2)} -
N_{d}^{(n - 2)}  + 1}\right)\left({N_{f}^{(n)}  - N_{d}^{(n)}  +
1}\right)} , \label{eq:corrpf}
\end{array}
\end{eqnarray}
where $P({g_n  = 1})$ is the probability that the corresponding
component $x_n$ exists in ${\bf y}_n$. As $N_{f}$ increases and/or
$N_{d}$ decreases in (\ref{eq:corrpf}), the correlation $E\left[
{y_{n - 2} y_n^* } \right]$ decreases (likewise for the
variance-normalized correlation). The same is true for the noise
correlation.

Fig. \ref{fig:corr} shows the example of correlation in the
innovation sequence before and after the refinement through
puncturing: $E[{\bf{x}}_{n - 2} {\bf{x}}_n^{H}]$ vs $E[{\bf{y}}_{n -
2} {\bf{y}}_n^{H}]$. The sequence ${\bf{y}}_n$ may have a smaller
number of observation samples, but its correlation is low as seen in
Fig.\ref{fig:corr} (b), which is useful to maintain the optimality
of the Kalman filter. The parameter $c$ controls trade-off : if $c$
is large, the number of observation samples increases, which can be
beneficial for ML estimation. However a large $c$ can feed biased
decision errors to the Kalman-based estimator.

Note that the actual puncturing process is not fully random as our
assumption made in (\ref{eq:corrpf}). However, the puncturing
happens irregularly, and an interesting observation we make is that
irregular puncturing activity become more pronounced in broken (bad)
packets. Once the decisions are incorrect, correlation between the
components of ${\bf{x}}_n$ appears, and puncturing becomes active.
In order to salvage a bad packet from biased errors, the puncturing
attempts to ``innovate" the sequence ${\bf{x}}_n$. Moreover, in high
SNR, random puncturing is not necessary to produce reliable
decisions, because the signal term itself in (\ref{eq:y}) (without
the noise and estimation error terms) is an innovation sequence.
Also, the puncturing process in this context can also be viewed as
an effort to prevent redundant information from circulating in the
iterative signal processing. We observe that although the puncturing
cannot completely remove the correlated errors, a significant
portion of the biased-errors gets eliminated before the channel
estimation step resumes.

\subsection{Kalman-Based Sequential Channel Estimation Algorithm with Punctured Innovation Sequence}\label{Kalman}

Once the punctured innovation sequence ${\mathbf{y}}_n$ is generated, a linear channel
estimator can be specified as
a matrix ${\mathbf{A}}$, that is, $\hat
{\mathbf{h}}=\mathbf{A}{\mathbf{y}}_n$. The Kalman estimator is now
derived as
\begin{eqnarray}
\hat {\mathbf{h}}_{n}  &=&
\widehat E[\mathbf{h}|\mathbf{y}_1,\mathbf{y}_2,...,\mathbf{y}_n]\nonumber \\
&=& \widehat
E[\mathbf{h}|\mathbf{y}_1,\mathbf{y}_2,...,\mathbf{y}_{n-1}]+\widehat
E[\mathbf{h}| \mathbf{y}_n]\nonumber \\
&=& \hat {\mathbf{h}}_{n-1} + \mathbf{A}_{n}\mathbf{y}_{n} ,
\label{eq:hhat}
\end{eqnarray}
where $\widehat E[\mathbf{a}|\mathbf{b}]$ denotes the optimal linear
estimator of  $\mathbf{a}$ given $\mathbf{b}$. To find the linear
estimator matrix $\mathbf{A}_n$, the orthogonality principle is
applied:
\begin{eqnarray}
\overline {\left( {{\bf{h}} - {\bf{A}}_n {\bf{y}}_n }
\right){\bf{y}}_n ^H }  = {\bf{0}} \nonumber \\
{\bf{A}}_n \overline {{\bf{y}}_n {\bf{y}}_n ^H}=\overline
{{\bf{h}}{\bf{y}}_n ^H },  \label{eq:OT}
\end{eqnarray}
where an overbar also indicates statistical expectation.
The right-hand-side of the last line in (\ref{eq:OT}) is
given by
\begin{eqnarray}
\overline  {{\bf{h}}{\bf{y}}_n ^H }=\underbrace {\overline {
({{\bf{h}}-\hat {\mathbf{h}}_{n-1}})({{\bf{h}}-\hat
{\mathbf{h}}_{n-1}})^H }}_{\buildrel \Delta \over = {\bf{P}}_{n - 1}
}\,\widetilde{\mathbf S}_n^H, \label{eq:hy}
\end{eqnarray}
where ${\bf{P}}_{n - 1}$ is defined as the channel estimation error
variance matrix, and the term $\overline {{\bf{y}}_n
{\bf{y}}_n ^H}$ in (\ref{eq:OT}) can be written as
\begin{eqnarray}
\overline {{\bf{y}}_n {\bf{y}}_n ^H}&=&\widetilde{\mathbf S}_n
\overline { ({{\bf{h}}-\hat {\mathbf{h}}_{n-1}})({{\bf{h}}-\hat
{\mathbf{h}}_{n-1}})^H }\,\widetilde{\mathbf S}_n^H  \nonumber \\
&&+ \underbrace { \overline{\mathbf{E}_n  {\bf{h}} {\bf{h}}^H
\mathbf{E}_n^H }}_{\buildrel \Delta \over = {{\bf{Q}}}_n}+ \mathcal
{N}_o {\bf{I}}_{N_d}. \label{eq:yy}
\end{eqnarray}
Now using (\ref{eq:OT}), (\ref{eq:hy}) and (\ref{eq:yy}), the matrix
${\mathbf A}_n$ is obtained as
\begin{eqnarray}
{\mathbf A}_n&=& \overline{{\bf{h}}{\bf{y}}_n ^H }(\overline
{{\bf{y}}_n
{\bf{y}}_n ^H})^{-1} \nonumber\\
&=&{\mathbf P}_{n-1} \widetilde{\mathbf S}_n^H  (\widetilde{\mathbf
S}_n {\mathbf P}_{n-1} \widetilde{\mathbf S}^H + {\mathbf Q}_n +
\mathcal {N}_o {\bf{I}}_{N_d} )^{-1} \nonumber \\&=& \left(
{\widetilde{\bf{S}}_n^H \left( {{\bf{Q}}_n {\bf{ + }}\mathcal {N}_o
{\bf{I}}_{N_d } } \right)^{ - 1} \widetilde{\bf{S}}_n {\bf{ + P}}_{n
- 1}^{ - 1} } \right)^{ - 1} \nonumber \\
&& \,\, \cdot \,\, \widetilde{\bf{S}}_n^H \left( {{\bf{Q}}_n {\bf{ +
}}\mathcal {N}_o {\bf{I}}_{N_d } } \right)^{ - 1}. \label{eq:A}
\end{eqnarray}
The next steps to complete the derivation process are to express
${\mathbf P}_{n-1}$ and ${\mathbf Q}_{n}$ in a recursive fashion.
Noticing $({\mathbf h}-\hat{\mathbf h}_n)={\mathbf h}-(\hat{\mathbf
h}_{n-1}+{\mathbf A}_{n}{\mathbf y}_n)$ from (\ref{eq:hhat}), the
channel estimation error variance at time $n$ can be rewritten as
\begin{eqnarray}
{\mathbf P}_{n} &=& \overline{\{{\mathbf h}-(\hat{\mathbf
h}_{n-1}+{\mathbf A}_{n}{\mathbf y}_n)\}\{{\mathbf h}-(\hat{\mathbf
h}_{n-1}+{\mathbf A}_{n}{\mathbf y}_n)\}^{H}} \nonumber  \\&=&
\overline{({\mathbf h}-\hat{\mathbf h}_{n-1})({\mathbf
h}-\hat{\mathbf h}_{n-1})^H}-\overline{({\mathbf h}-\hat{\mathbf
h}_{n-1}){\mathbf y}_n^H}{\mathbf A}_n^H \nonumber \\ && -{\mathbf
A}_n\overline{{\mathbf y}_n ({\mathbf h}-\hat{\mathbf
h}_{n-1})^H}+{\mathbf A}_n\overline{{\mathbf y}_n{\mathbf
y}_n^H}{\mathbf A}_n^H \nonumber  \\
&=&{\mathbf P}_{n-1}-{\mathbf A}_n\widetilde{\mathbf S}_n{\mathbf
P}_{n-1}^H \nonumber \\ &=& ({\mathbf I}_{N_t}-{\mathbf
A}_n\widetilde{\mathbf S}_n){\mathbf P}_{n-1}, \label{eq:P}
\end{eqnarray}
where we utilized the relation $\overline{{\mathbf y}_n{\mathbf
y}_n^H}{\mathbf A}_n^H=\widetilde{\mathbf S}_n{\mathbf P}_{n-1}^H $
which is obvious from (\ref{eq:OT}) and (\ref{eq:hy}). Also note
${\mathbf P}_{n}$ is a symmetric matrix of which pivot has
non-negative real values.

Finally, ${\mathbf Q}_n$ needs to be found. The symbol decision
error variance $\sigma_s^2=E[|s-\tilde s|^2]$ can be found by using
the extrinsic probabilities (i.e. $\sigma_s^2 =\sum\nolimits_{\tiny{s_i
\in {\mathcal {A}}}} {|s_i - \tilde s|^2 P(s_i )}$). Under the
reasonable assumption of
$\overline{(s_j-\tilde{s}_j)(s_i-\tilde{s}_i)^*}=0$ when $i \ne j$,
the $N_d \times N_d$ diagonal matrix ${\mathbf Q}_n$ is given as
\begin{eqnarray}
\overline{{\mathbf E}_n{\mathbf {hh}}^H{\mathbf
E}_n^H}&=&{\rm{diag}}\left[ {\sum\nolimits_{t = 1}^{N_t }
{\rho^{(t)} } \sigma _s^2 (n,0,t)},\,...,  \right. \nonumber \\
&& \left.{\sum\nolimits_{t = 1}^{N_t } {\rho ^{(t)} } \sigma _s^2
(n,N_d  - 1,t)} \right]_{N_d \times N_d} ,
\end{eqnarray}
where ${\rho^{(t)}} \buildrel \Delta \over = |h^{(t)}|^2 $. However,
finding ${\rho^{(t)}}$ is a bit tricky as the channel state
information is unknown to the receiver. The channel correlation
matrix $ {\mathbf {hh}}^H$, on the other hand, can be found from
$\overline {{\bf{hh}}^H }=\overline{\{({\mathbf h}-{\hat{\mathbf
h}}_n)+{\hat{\mathbf h}}_n\}\{({\mathbf h}-{\hat{\mathbf
h}}_n)+{\hat{\mathbf h}}_n\}^H}$, which reduces to ${\mathbf
P}_n+{{\hat{\mathbf h}}_n{\hat{\mathbf {h}}}_n^H}$. Utilizing this
expression, we can write
\begin{eqnarray}
&{\mathbf Q}_{n}& =  \overline {{\bf{E}}_n \left( {{\bf P}_n  + {\bf
\hat h\hat h}_n^H } \right){\bf{E}}_n^H } \nonumber \\ &=&
{\rm{diag}}\left[{\sum\limits_{t = 1}^{N_t } {\left( {p_n (t,t) +
|\hat{h}^{(t)}_{n-1} |^2 } \right)\sigma_s^2 (n,0,t) } },...,\right.\nonumber  \\
&&\left. {\sum\limits_{t = 1}^{N_t } {\left( {p_n (t,t) +
\hat{h}^{(t)}_{n-1} |^2 } \right)\sigma_s^2 (n,N_d-1,t) } } \right],
\label{eq:Q}
\end{eqnarray}
where $\hat h_t[n-1]$ is from the previous estimate $\hat
{\mathbf{h}}_{n-1}$, $p_n{(t,t)}$ is the $t^{th}$ diagonal element
of ${\mathbf P}_{n-1}$, and $\sigma_s^2 (n,j,t)$ is the decision
error variance of the $(j,t)$ element of $\widetilde {\mathbf S}_n$.

Putting it all together, for the receive antenna $r$, the proposed
estimator is summarized as a set of equations :
\begin{eqnarray}
{\mathbf Q}_{n}^{(r)} &=& {\rm{diag}}\left[{ {\sum\limits_{t =
1}^{N_t } {\left( {p_n (t,t) + |\hat{h}^{(t)}_{n-1} |^2 }
\right)\sigma_s^2 (n,0,t) } },\cdot  \cdot  \cdot, } \right.\nonumber  \\
&& \left. { {\sum\limits_{t = 1}^{N_t } {\left( {p_n (t,t) +
|\hat{h}^{(t)}_{n-1} |^2 } \right)\sigma_s^2 (n,N_d-1,t) } } }
\right]_{N_d\times N_d}\label{eq:MESTI}\\ {\mathbf
A}_n^{(r)}&=&\left( {\widetilde{\bf{S}}_n^H \left( {{\bf{Q}}_n^{(r)}
{\bf{ + }}\mathcal {N}_o {\bf{I}}_{N_d } } \right)^{ - 1}
\widetilde{\bf{S}}_n {\bf{ + P}}_{n - 1}^{ {(r)}- 1} } \right)^{ -
1} \nonumber \\ && \widetilde{\bf{S}}_n^H \left( {{\bf{Q}}_n^{(r)}
{\bf{ + }}\mathcal {N}_o {\bf{I}}_{N_d } } \right)^{ -
1} \label{eq:estA} \\
{\mathbf P}_{n}^{(r)}&=&({\mathbf I}_{N_t}-{\mathbf
A}_n^{(r)}\widetilde{\mathbf S}_n){\mathbf P}_{n-1}^{(r)}\label{eq:estP} \\
\hat {\mathbf{h}}_{n}^{(r)}&=&\hat {\mathbf{h}}_{n-1} +
\mathbf{A}_{n}\mathbf{y}_{n}\label{eq:hest},
\end{eqnarray}
where $\hat {\mathbf{h}}_{-1}$ corresponding to the initial time
$n=0$ can be given by an initial channel estimator based on the use of known preambles.
Also the initial matrix ${\mathbf
P}_{-1}^{(r)}$ can be derived from the MMSE analysis~\cite{kay} as
${\mathbf P}_{-1}^{(r)}=\rm{diag}[ |\hat
{h}_{-1}^{(r,t)}|^2/({\gamma |\hat {h}_{-1}^{(r,t)}|^2+1})]$ for
$t=1,..,N_t$ where $\gamma=E_s/(N_t \mathcal {N}_o)$. We note that
the channel estimation algorithm summarized in
(\ref{eq:MESTI})-(\ref{eq:hest}) takes into account the quality of
the soft decisions that are generated at various stages in the
pipeline for a given processing time $n$. When $t$=1, the resulting
algorithm becomes similar to the one presented in \cite{Song04} for
the inter-symbol interference channel, as the gist of the algorithm
of \cite{Song04} is in incorporating the quality of the soft
decisions as part of effective noise in the Kalman sequential
updating process. The difference, however, is that in our algorithm,
we do not assume that the operation of $\mathbf{z}_n-
\widetilde{\mathbf S}_n \hat {\mathbf{h}}_{n-1}$ makes the
observation sequence automatically white, which, as argued above,
would be faulty. Also, in our algorithm, varying qualities of the
decisions generated from different processing modules at \emph{a
given time} are taken into account in the update process. More
specifically, the effective noise covariance matrix of
(\ref{eq:MESTI}) is a function not only of $n$ but also of $N_d$
which itself is a growing function of $n$ initially (up to
$2N_{itr}$).

\subsection{Noise Variance Update for the Soft Detectors}\label{noisevar}

A Kalman-based estimation algorithm, as
the one proposed here, has the advantage (compared with, e.g.,
EM-like algorithms) that the channel estimation error variance is
available for free and it is continually updated as a part of the
recursive process. Realizing that the channel estimation error
variance is a reasonable measure of how accurate the channel
estimate is, this information somehow should play a beneficial role in the
detection (or demapping) process.

As the first step in utilizing the available channel estimation error
variance, the observation equation of
(\ref{eq:z}) is recast with the channel estimation error shown explicitly:
\begin{equation}
{\mathbf{z}}^{(k)}_n={ \mathbf{\widehat H}} {\mathbf{s}}_n^{(k)}  +
\underbrace {{ \left( \mathbf{H}-\widehat {\mathbf{H}}_n\right)}
\mathbf{s}_n^{(k)}}_{ \buildrel \Delta \over = {\bf{a}}_n^{(k)}}
+\mathbf{n}^{(k)}_n, \ \label{eq:hz}
\end{equation}
where superscript $k$ points to a specific demapper out of the $N_{itr}$
demappers operating in the pipeline stages ($k=1,..,N_{itr}$). Accordingly,
$\mathbf{s}_n^{(k)}$ here corresponds to each odd row of
${\bf{S}}_n$ in (\ref{eq:Mz}). For the $k^{th}$ demapper in the
pipeline, the noise variance is updated to include the channel
estimation error:
\begin{eqnarray}
\widehat{\mathbf{N}}_{o}^{(k)}[n] &=& {E\left\{ {\left\|
{{\bf{s}}^{(k)}_n ({\bf{H}} - \widehat{\bf{H}}_n ) + {\rm \mathbf
{n}}^{(k)} } \right\|^2 } \right\}} \nonumber \\ &=&  {{\bf
Cov}({\bf{a}}_n^{(k)},{\bf{a}}_n^{(k)})  + \mathcal N_o
{\bf{I}}_{N_r}}, \label{eq:upvar}
\end{eqnarray}
where $\left\|\,\right\|$ indicates vector norm operation. The
${N_r\times N_r}$ covariance matrix ${{\bf
{Cov}}({\bf{a}}_n^{(k)},{\bf{a}}_n^{(k)})}$ can be obtained (with an
understanding we are focusing on the $k^{th}$ demapper in the
pipeline at time $n$, drop the indices $k$ and $n$ to simplify
notation) as
\begin{eqnarray}
\begin{array} {l}
{{\bf Cov}({\bf{a}},{\bf{a}})}= {E\left\{ { {({\bf{H}} -
\widehat{\bf{H}} )^H }{\bf{s}}^{H} {{\bf{s}} ({\bf{H}}
-\widehat{\bf{H}} ) }  } \right\}}  \\
\approx {\rm{diag}}\left[ {\sum\limits_{t = 1}^{N_t } {p^{(1)}
(t,t)|\widetilde s^{(t)} |^2 ,..,\sum\limits_{t = 1}^{N_t } {p^{(N_r
)} (t,t)|\widetilde s^{(t)} |^2 } } } \right], \label{eq:Cov}
\end{array}
\end{eqnarray}
where the approximation is due to the assumption that channel
estimation errors and transmitted symbols are independent and that $
E[{\bf{s}}_n^H {\bf{s}}_n ] \approx E[\widetilde{\bf{s}}_n^H
\widetilde{\bf{s}}_n ]$. Note that the updated noise variance is specified in
matrix form because different RX antennas are subject to different
channel estimate errors in the Kalman estimator. This is the same as saying
each RX antenna is subject to a different amount of observation noise.
Therefore, the demapper algorithm must properly be optimized for the given
equivalent noise covariance matrix.

The demappers in the pipeline utilize (\ref{eq:hz}). An $M$-QAM symbol vector transmitted
from $N_t$-TX streams is demapped to one binary vector ${\bf{b}} =
[b_0 ,b_1 , \cdots ,b_{QN_t - 1} ]^T$. Using the updated noise
variance, the likelihood function of the MIMO demapper is given as
\begin{eqnarray}
\begin{array}{l}
\displaystyle P({\bf{z}}|{\bf{s}}) = \prod\limits_{r = 1}^{N_r }
{\frac{1}{{\sqrt {2\pi \widehat {\mathcal N}_o^{(r)} } }}\exp \left(
{ - \frac{{|z^{(r)}  - \widehat{\bf{h}}^{(r)} {\bf{s}}|^2
}}{{\widehat
{\mathcal N}_o^{(r)} }}} \right)}  \\
\displaystyle =\frac{1}{{\left( {\sqrt {2\pi } } \right)^{N_r }
\prod\limits_{r = 1}^{N_r } {\widehat {\mathcal N}_o^{(r)} } }}\exp
\left( { - \sum\limits_{r = 1}^{N_r } {\frac{{|z^{(r)}  -
\widehat{\bf{h}}^{(r)} {\bf{s}}|^2 }}{{\widehat {\mathcal N}_o^{(r)}
}}} } \right),
\end{array}
\end{eqnarray}
where ${\widehat{\mathcal N}_o^{(r)}}$ is the noise variance
corresponding to $z^{(r)}$, that is, the $(r,r)$ diagonal element of
matrix $\widehat{\mathbf{N}}_{o}$. The $k^{th}$ soft MAP demapper in
the pipeline directly gives out the posteriori LLR output $L_P$:
\begin{eqnarray}
\begin{array}{l}
\displaystyle L_P \left({b_i } \right) = \ln \frac{{P\left( {b_i  =
1|{\bf{z}}
} \right)}}{{P\left( {b_i  = 0|{\bf{z}} } \right)}}  \\
\displaystyle = \ln \frac{{\sum\limits_{{\bf{s}} \in {\mathcal
{A}^{Nt}} |b_i = 1} {P\left( {{\bf{z}}|{\bf{s}}}
\right)\prod\limits_{j \ne i} {P\left( {b_j } \right)} }
}}{{\sum\limits_{{\bf{s}} \in {\mathcal {A}}^{Nt} |b_i  = 0}
{P\left( {{\bf{z}}|{\bf{s}}} \right)\prod\limits_{j \ne i} {P\left(
{b_j } \right)} } }} +  \ln \frac{{P\left( {b_i  = 1}
\right)}}{{P\left( {b_i  = 0} \right)}}, \label{eq:MAPdet}
\end{array}
\end{eqnarray}
where $i=0,...,{QN_t - 1}$ for the individual bits in the transmitted
symbol vector.

The MMSE demapper solution can also be derived from the modified observation equation
(\ref{eq:hz}). The MMSE demapper can be shown to yield
\begin{eqnarray}
\widehat{\bf{s}} &=& E[{\bf{s}}] + {\bf{ \Sigma }}_{\bf{s}} {
\bf{\widehat H}}^{H} \left( {\bf{\widehat H\Sigma }}_{\bf{s}} {
\bf{\widehat H}}^{H} + {{\bf Cov}(\bf{a},\bf{a})} + {\mathcal N}_o
{\bf{I}} \right)^{ - 1} \nonumber \\ && \cdot \left( {{{{\bf z} -
{\bf \widehat H} E[{\bf{s}}] }} } \right),  \label{eq:MMSEdet}
\end{eqnarray}
where  $E[{\bf{s}}]$ is a mean-symbol vector based on the $a priori$
probabilities, and ${\bf{\Sigma }}_{\bf{s}}$ is given as
${\rm{diag}} [ {\sigma _{s_0 }^2 , \ldots ,\sigma _{s_{N_t  - 1} }^2
}]$.

\section{Mean Squared Error (MSE) Analysis}\label{MSE}

In the MSE analysis, we try to understand 1) the impact of biased
soft decision errors, and 2) when the soft decision error is
unbiased, the performance impact of mismatching the soft decision
error variance in the estimation channel process. Through the MSE
analysis, we also investigate the impacts of the number and quality
of decisions used in the estimation process.

The soft decisions fed back from the detectors and decoders are
assumed to have potential errors and are written as
\begin{eqnarray}
\tilde s[d,n] = s[d,n]+e[d,n],
\end{eqnarray}
where $d=0,..,N_d-1$. As discussed in Section \ref{Kalman},
the fed-back soft-decisions may contain
biased decision errors. So the decision error $e_d$ is modeled as
\begin{eqnarray}
e[d,n] = m[n] + q[d,n],
\end{eqnarray}
where $m$ is a non-zero-mean random variable, and $q$ is a white Gaussian
noise with zero mean and variance $\sigma_q^2$. Also, denote
$\sigma_s^2=E[|e|^2]$. For both biased and unbiased cases, assume
that the total decision error power $\sigma_s^2$ is identical. Also
assume correlations of the bias mean with the symbol as well as with
the channel are zero (i.e. $E[s\,m]=0$ and $E[h\,m]=0$).

The proposed estimator is designed based on the linear MMSE (LMMSE)
criterion. For the MISO communication channel of (\ref{eq:Mz}), the
LMMSE estimator is expressed as
\begin{eqnarray}
\hat{\bf{h}}_n^{(r)}  &=& {\bf{A}}_n^{(r)} {\bf{z}}_n^{(r)}  \\
{\bf{A}}_n^{(r)} &=& {\bf{R}}_h^{(r)} \widetilde{\bf{S}}_n^H \left(
{\widetilde{\bf{S}}_n {\bf{R}}_h^{(r)} \widetilde{\bf{S}}_n^H +
{\bf{V}}_n^{(r)}  } \right)^{ - 1} \nonumber \\&=&\underbrace
{\left( {\widetilde{\bf{S}}_n^H \widetilde{\bf{S}}_n  + v_n^{(r)}
{\bf{R}}_h^{(r){-1}} } \right)^{ - 1}}_{ \buildrel \Delta \over =
{\bf{\Psi }}^{-1}} \widetilde{\bf{S}}_n^H, \label{eq:correct}
\end{eqnarray}
where ${\bf{V}}_n^{(r)}=v_n^{(r)}{\bf{I}}_{N_d }$ with $
v^{(r)}_n=(\sigma_{s,n}^2 {\sum\nolimits_{t = 1}^{N_t }
{\rho^{(r,t)} } } + \mathcal {N}_o)$, and ${\bf{R}}_h^{(r){-1}}  =
E\left\{ {\bf{h}}^{(r)} {\bf{h}}^{(r){H}} \right\}$. Also, denote
${\bf{\Psi }} \buildrel \Delta \over = \widetilde{\bf{S}}_n^H
\widetilde{\bf{S}}_n + v_n {\bf{R}}_h^{(r){-1}}$. The estimator of
(\ref{eq:correct}) is optimum under unbiased decision errors
($m=0$), and the minimum estimation error variance of the MIMO LMMSE
estimator is obtained as
\begin{eqnarray}
\varepsilon_{unbiased}^2[n] &=& \sum\limits_{r=1}^{N_r} {
E\left\{\|{\mathbf
h}^{(r)}- \hat{{\mathbf{h}}}_n^{(r)}\|^2\right\}}   \\
&=& \sum\limits_{r=1}^{N_r} { {\bf tr} \left\{ { \left(
{\widetilde{\bf{S}}_n^H {\bf{V}}_n^{(r)-1} \widetilde{\bf{S}}_n +
{\bf{R}}_h^{(r){-1}}  } \right)^{ - 1}} \right\} } \nonumber
\label{eq:epsilon},
\end{eqnarray}
where $\varepsilon_{unbiased}^2$ is the estimation error variance of
the optimum MMSE estimator (\ref{eq:correct}). As $N_d$ increases, it is reasonable to write $ {
\widetilde{\bf{S}}^H \widetilde{\bf{S}}}= N_d
E\{\widetilde{\bf{S}}^H \widetilde{\bf{S}}\}
=N_d(E_s+\sigma_s^2){\bf{I}}_{N_t}$. Accordingly, we have
\begin{eqnarray}
\varepsilon _{opt}^2 =\varepsilon _{unbiased}^2  = \sum\limits_{r =
1}^{N_r } {\sum\limits_{t = 1}^{N_t } {\frac{1}{{N_d (E_s  + \sigma
_s^2 )/v^{(r)}  + 1/\rho ^{(r,t)} }}}}  \label{eq:optep}.
\end{eqnarray}
Meanwhile, when $m \ne 0 $ with the same decision error power
$\sigma_s^2$, the MSE with the biased decision error is calculated
as
\begin{eqnarray}
\varepsilon _{biased}^2  = \sum\limits_{r = 1}^{N_r } {E\left\{ {||
{{\bf h}^{(r)}  - \widehat{\bf{h}}_{biased}^{(r)} } ||^2 } \right\}}
\label{eq:epsilon},
\end{eqnarray}
where $\widehat{\bf{h}}_{biased}={\bf{A}}_{biased} {\bf{z}}$ and
${\bf{A}}_{biased}=({\widetilde{\bf{S}}_{biased}^H
\widetilde{\bf{S}}_{biased} + v{\bf{R}}_h^{ - 1} })^{ - 1}
\widetilde{\bf{S}}_{biased}^H$ that utilizes soft-decisions with
correlated error. Note that the correlation matrix of decision
errors is $ E[{\bf{E}}^H {\bf{E}}] = N_d \sigma _s^2 \,{\bf{I}}_{Nt}
+ N_d {\bf{\Phi }}_{N_t }$, where ${\bf{\Phi}}_{N_t}$ is a matrix
with all diagonal elements set to zeros and all non-diagonal
elements to $|m|^2$. Assuming a very large $N_d$ and applying a
matrix inversion lemma $ ({\bf{X}} + {\bf{Y}})^{ - 1} = {\bf{X}}^{ -
1} - {\bf{X}}^{ - 1} ({\bf{X}}^{ - 1}  + {\bf{Y}}^{ - 1} ){\bf{X}}^{
- 1}$, we can write
\begin{eqnarray}
\begin{array} {l}
\left( {\widetilde{\bf{S}}_{biased}^H \widetilde{\bf{S}}_{biased} +
v_n {\bf{R}}_h^{ - 1} } \right)^{ - 1}
\\= (\underbrace {N_d \left( {E_s  + \sigma _s^2 } \right){\bf{I}}_{Nt}
+ v_n{\bf{R}}_h^{ - 1} }_{ = {\bf{\Psi }}} + N_d {\bf{\Phi }}_{Nt} )^{ - 1}  \\
= {\bf{\Psi }}_{}^{ - 1}  - \underbrace {{\bf{\Psi }}_{}^{ - 1}
\left( {{\textstyle{1 \over {N_d }}}{\bf{\Phi }}_{N_t }^{ - 1}  +
{\bf{\Psi }}_{}^{ - 1} } \right)^{ - 1} {\bf{\Psi }}_{}^{ - 1} }_{ =
{\bf{\Lambda }}}. \label{eq:invle}
\end{array}
\end{eqnarray}
Using (\ref{eq:invle}) and the facts $\,{\bf{tr}}\left( {{\bf{\Psi
}}^{ - 1} {\bf{\Phi }}_{Nt} {\bf{R}}_h } \right) = 0$ and
${\bf{tr}}\left( {{\bf{\Phi }}_{Nt} {\bf{R}}_h } \right) = 0$, the
MSE of the LMMSE estimator suffering from correlated decision errors
is finally expressed as
\begin{eqnarray}
\varepsilon _{biased}^2 &=& \sum\limits_{r = 1}^{N_r }
{{\bf{tr}}\left\{ {{\bf{R}}_h^{(r)}  -
\widehat{\bf{h}}_{unbiased}^{(r)} {\bf{h}}^{(r)H} } \right\}  }
\nonumber \\&& + {\bf{tr}}\left\{ {\left( {N_d \left( {E_s  + \sigma
_s^2 } \right){\bf{\Lambda }}^{(r)} {\bf{R}}_h^{(r)} } \right)}
\right\}  \\&=& \varepsilon _{unbiased}^2  + \sum\limits_{r =
1}^{N_r } {{\bf{tr}}\left\{ {\left( {N_d \left( {E_s  + \sigma _s^2
} \right){\bf{\Lambda }}^{(r)} {\bf{R}}_h^{(r)} } \right)}
\right\}}. \nonumber \label{eq:epb}
\end{eqnarray}
Note that ${{\bf{\Lambda }} {\bf{R}}_h}$ is a semi-positive definite
matrix, and therefore $\varepsilon _{biased}^2 > \varepsilon
_{unbiased}^2$ when $|m|\ne0$. This confirms the loss due to
correlated decision errors, even if the error power is the same.
\begin{figure}
\centering
\includegraphics[width=3.5in, height=2.9in]{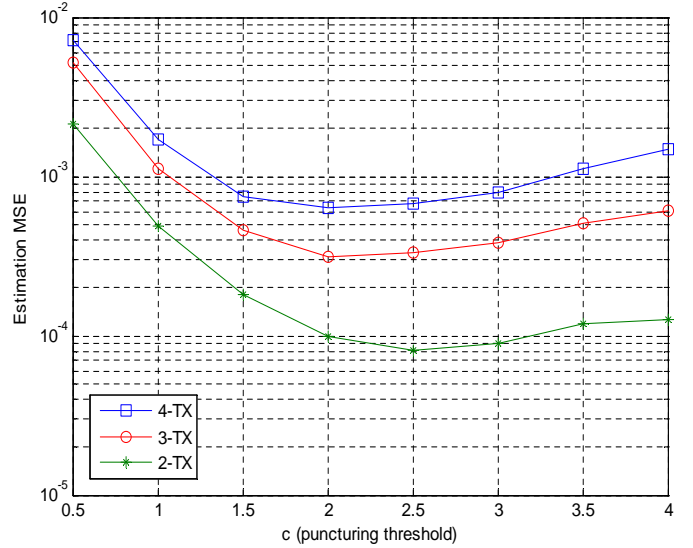}
\caption{Threshold parameter $c$ optimization} \label{fig:cpara}
\end{figure}
In effectively whitening the correlated decision error, the constant
$c$ is a crucial parameter that determines the number of selected
symbols $N_d$ and thus controls the trade-off between the
observation sample size and the amount of error correlation in the
channel estimator. The existence of an optimum value for $c$ is also
shown through the MSE simulation results of Fig.~\ref{fig:cpara}.
Based on Fig.~\ref{fig:cpara}, we set $c=2$ for the $3\times 3$ and
$4\times 4$ SM-MIMO-OFDM systems, and $c=2.5$ for the $2\times 2$
SM-MIMO-OFDM system.

Even with unbiased decision errors, the LMMSE estimator
suffers performance degradation when the noise variance is
mismatched. Let us quantify the MSE penalty associated with not
accounting for the uncertainty inherent in the soft decisions in the
form of increased noise variance. The LMMSE estimator failing to
consider the soft decision error can be described as
\begin{eqnarray}
\hat{\bf{w}}_n^{(r)}  &=& {\bf{W}}_n^{(r)} {\bf{z}}_n^{(r)}  \\
{\bf{W}}_n^{(r)} &=& {\bf{R}}_h^{(r)} \widetilde{\bf{S}}_n^H \left(
{\widetilde{\bf{S}}_n {\bf{R}}_h^{(r)} \widetilde{\bf{S}}_n^H +
\mathcal {N}_o{\bf{I}}_{N_d} } \right)^{ - 1} \nonumber
\\&=&\left({\widetilde{\bf{S}}_n^H  \widetilde{\bf{S}}_n
+\mathcal {N}_o{\bf{R}}_{h}^{(r) - 1} } \right)^{ - 1}
\widetilde{\bf{S}}_n^H.
 \label{eq:wrong}
\end{eqnarray}
\begin{figure}
\centering
\includegraphics[width=3.5in, height=2.9in]{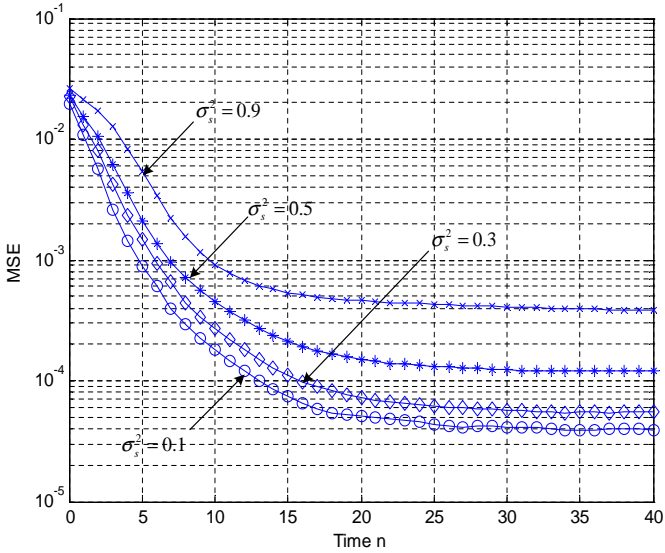}
\caption{Open-loop channel estimation MSE for different values of
$\sigma_s^2$ ($N_d=12$)} \label{fig:MSE_sig}
\end{figure}
Utilizing (\ref{eq:optep}) and denoting $ {\bf\Delta}_n^{(r)}
\buildrel \Delta \over = \hat{\bf{h}}_n^{(r)} -
\hat{\bf{w}}_n^{(r)}$, the estimation error variance
$\varepsilon_{w}^2$ of the estimator (\ref{eq:wrong}) can be shown
to be
\begin{eqnarray}
\varepsilon_{w}^{2} [n] &=& \sum\limits_{r=1}^{N_r} {
E\left\{\|{\mathbf h}_n^{(r)}-\hat{\mathbf{w}}_n^{(r)}\|^2\right\}}
\nonumber\\&=& \sum\limits_{r = 1}^{N_r } {{\bf{tr}} {E\left\{ {\|
{{\bf{h}}_n^{(r)}  - \left( {\hat{\bf{h}}_n^{(r)} - {\bf{\Delta
}}_n^{(r)} } \right)} \|^2} \right\}} }  \nonumber \\ &=&
\varepsilon _{opt}^2[n] + \sum\limits_{r = 1}^{N_r }
{\bf{tr}}\left\{E\left\{ {{\bf{h}}_n^{(r)} \hat{\bf{h}}_n^{(r)H} -
\hat{\bf{w}}_n^{(r)} \hat{\bf{w}}_n^{(r)H} }  \right. \right.
\nonumber \\  && \left. \left. +
\hat{\bf{w}}_n^{(r)}{\bf{h}}_n^{(r)H}-{\bf{h}}_n^{(r)}
\hat{\bf{w}}_n^{(r)H} \right\} \right\} \label{eq:errorw}.
\end{eqnarray}
To simplify notation, the indices $r$ and $n$ are temporally
dropped. As the number of iteration increases, the matrix inversions
in (\ref{eq:correct}) and (\ref{eq:wrong}) can be simplified as
$\left( {\widetilde{\bf{S}}^H  \widetilde{\bf{S}} + {v}
{\bf{R}}_{h}^{ - 1} } \right)^{ - 1} = {\rm{diag}}[ \,\, {{\rho _1
}}/ {{ \left(\rho _1 N_d ( {E_s  + \sigma _{s}^2 } ) +
v\right)}}$,...,$\\{{\rho_{N_t} }}/{{ \left(\rho_{N_t} N_d ( {E_s +
\sigma _{s}^2 } ) + v\right) }} \,\, ] \,\,$ and $ \left(
{\widetilde{\bf{S}}^H \widetilde{\bf{S}} +\mathcal {N}_o
{\bf{R}}_{h}^{ - 1} } \right)^{ - 1} = {\rm{diag}}[{{ \rho _1 }}/{{
\left(\rho_1 N_d ( {E_s + \sigma _{s}^2 } ) + \mathcal {N}_o
\right)}}$,...,${{\rho_{N_t} }}/\left(\rho_{N_t} N_d ( {E_s + \sigma
_{s}^2 } \right)$  $\\+ \mathcal {N}_o )]$, where the subscript for
$\rho$ for the time being indicates the TX antenna. Also, noting
$E\{{\hat {\bf{h}}( {{\bf{h}} - \hat {\bf{h}}})^H}\}=0$ by the
orthogonality principle, it can be shown that
\begin{eqnarray}
\begin{array}{l}
\,\, E\{ {{\bf{h}}\hat{\bf{h}}^H }\} = {\bf{R}}_h {\bf{S}}^H
\widetilde{\bf{S}}\left( {\widetilde{\bf{S}}^H \widetilde{\bf{S}} +
{v} {\bf{R}}_{h}^{ - 1} } \right)^{ - H} \\ \,\,= {\rm{diag}}\left[
\frac{{\rho _1^2 N_d {E_s}}}{{\rho _1 N_d \left( {E_s  + \sigma _s^2
} \right) + v}} ,..., \frac{{\rho _{N_t }^2 N_d{E_s}}}{{\rho _{N_t }
N_d \left( {E_s  + \sigma _s^2 } \right) + v}} \right]_{N_t \times
N_t}.
\end{array} \label{eq:wh}
\end{eqnarray}
\begin{figure}
\centering
\includegraphics[width=3.5in, height=2.9in]{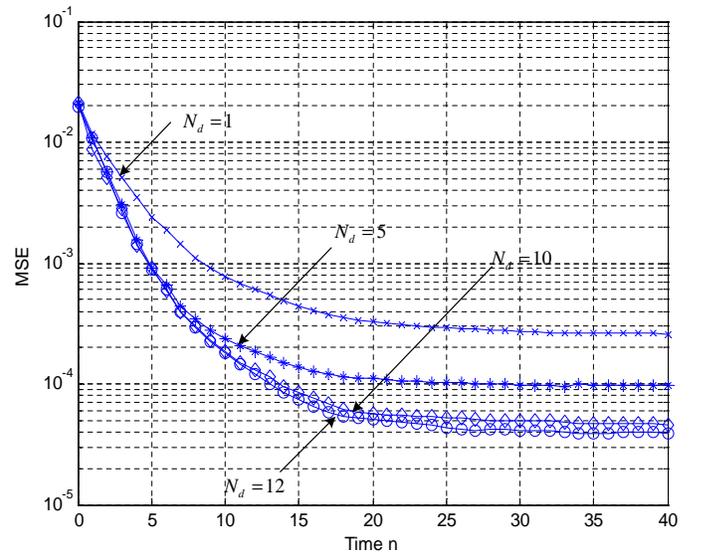}
\caption{Open-loop channel estimation MSE depending on different
values of $N_d$ ($\sigma_s^2=0.1$)} \label{fig:MSE_Nd}
\end{figure}
We also write
\begin{eqnarray}
\begin{array}{l}
E\{\hat{\bf{w}} \hat{\bf{w}}^H \} \\ = \left( {\widetilde{\bf{S}}^H
\widetilde{\bf{S}} +\mathcal {N}_o {\bf{R}}_{h}^{ - 1} } \right)^{ -
1} \widetilde{\bf{S}}^H \left( {{\bf{S R}}_h {\bf{S}}^H + \mathcal
{N}_o{\bf{I}}_{N_d}} \right)
 \\ \,\,\,\,\,\,\,\,\, \cdot \, \widetilde{\bf{S}}\left( {\widetilde{\bf{S}}^H
\widetilde{\bf{S}} +\mathcal
{N}_o {\bf{R}}_{h}^{ - 1} } \right)^{ - H}  \\
= {\rm{diag}}\left[ \frac{{\rho _1^2 N_d \left( {\rho _1 N_d E_s ^2
+ {\rho _{_{\Sigma}}}  \sigma _s^2 E_s  + \mathcal {N}_o \left( {E_s
+ \sigma _s^2 } \right)} \right)}}{{\left( {\rho _1 N_d \left( {E_s
+ \sigma _s^2 } \right) + \mathcal {N}_o } \right)^2 }},...,\right.
\\\,\,\,\,\,\,\,\,\,  \left.
\frac{{\rho _{N_t }^2 N_d \left( {\rho _{N_t } N_d^{} E_s ^2  +
{\rho _{_{\Sigma}}} \sigma _s^2 E_s + \mathcal {N}_o \left( {E_s  +
\sigma _s^2 } \right)} \right)}}{{\left( {\rho _{N_t } N_d \left(
{E_s + \sigma _s^2 } \right) + \mathcal {N}_o } \right)^2 }}
\right]_{N_t \times N_t},
\end{array}\label{eq:ww}
\end{eqnarray}
where ${\rho_{_{\Sigma}}}={\sum\nolimits_{t = 1}^{N_t } {\rho _t }
}$.
Finally, substituting (\ref{eq:wh}) and (\ref{eq:ww}) in (\ref{eq:errorw})
and also noting $E\{ \hat{\bf{w}}{\bf{h}}^H \} = E\{ {\bf{h}}\hat{\bf{w}}^H
\}$, the MSE convergence behavior of the estimator
(\ref{eq:errorw}) can be shown to be
\begin{eqnarray}
\begin{array}{l}
\displaystyle \mathop {\lim }\limits_{N_d  \to \infty } \varepsilon
_w^2 [n] \\ \displaystyle = \varepsilon _{opt}^2[n] + \sum\limits_{r
= 1}^{N_r } { \frac{{\rho ^{(r)}_{\Sigma} E_s }}{{E_s  + \sigma
_s^2[n] }} \left( {1 - \frac{{E_s }}{{E_s + \sigma _s^2[n] }} }
\right)}, \label{eq:MSE}
\end{array}
\end{eqnarray}
from which it is easy to see that the mismatched MSE is an increasing function of
the soft decision error variance $\sigma _s^2$.

To develop insights into the performance sensitivity off the
sequentially updated channel estimator against the variations of the
parameters $\sigma_s^2$ and $N_d$, we resort to an open-loop
investigation. For this, the decision-feedback channel estimator is
modified in such a way that unbiased soft decisions with various
$\sigma_s^2$ and $N_d$ combinations are artificially generated for
the channel estimator. A 7-iteration IDD receiver for the $2 \times
2$ 16-QAM MIMO-OFDM system is used for this test, but instead of
using actual feedback from the demappers and decoders, artificially
generated soft-decisions are provided to the channel estimator of
(\ref{eq:MESTI})-(\ref{eq:hest}).

Fig.~\ref{fig:MSE_sig} and Fig.~\ref{fig:MSE_Nd} show the MSE
performance depending on the decision quality $\sigma_s^2$ and the
number of feedback decisions $N_d$ with an assumption of
uncorrelated feedback decisions. The signal power is fixed at
$E_s=1$ and the channel SNR at 14 dB. With $N_d=12$, the packet
error rate (PER) due to imperfect CSI became negligible when $\sigma
_s^2 \approx 0.1$. In Fig.~\ref{fig:MSE_Nd}, it is seen that
reducing the number of feedback decisions, $N_d$, while fixing the
decision quality causes the MSE to increase.

\section{Performance Evaluation}\label{PE}

The proposed algorithm is investigated through an extrinsic
information transfer (EXIT) chart analysis and packet error rate
(PER) simulation. Performances are evaluated for $2\times 2$,
$3\times 3$ and $4\times 4$ 16-QAM SM-MIMO-OFDM systems. The
transmitter sends a packet with $1000$ bytes of information. The
SISO MAP-demapper is used for the $2\times 2$ SM-MIMO-OFDM system,
whereas the SISO MMSE-demapper is used for the $3\times 3$ and
$4\times 4$ SM-MIMO-OFDM system \cite{Tuchler02_1} due to
complexity. A rate-$1/2$ convolutional code is used with generator
polynomials $g_o  = 133_8$ and $g_1  = 171_8$, complying with the
IEEE 802.11n specifications \cite{IEEE11n}. The SOVA is used for
decoding. The MIMO multi-path channel is modeled with an
exponentially-decaying power profile with $T_{rms}=50ns$
uncorrelated across the TX-RX links established.

\subsection{EXIT and PER Performance Comparisons}

\begin{figure}
\centering
\includegraphics[width=3.5in, height=2.8in]{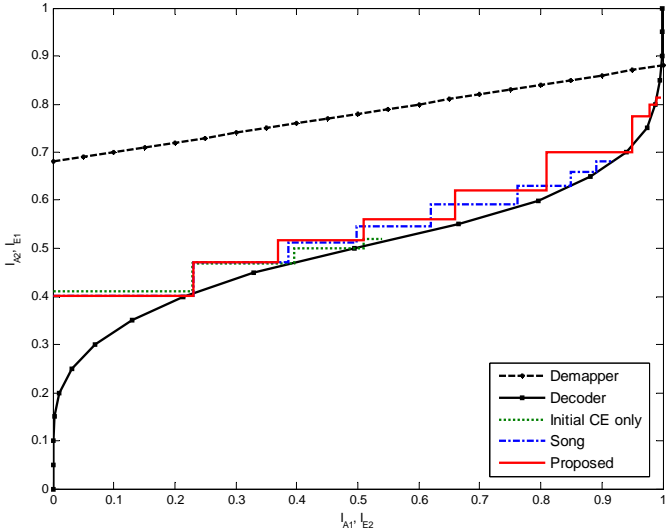}
\caption{EXIT charts for the $2\times 2$ SM-MIMO-OFDM turbo receiver
at SNR=14 dB} \label{fig:exit}
\end{figure}
The EXIT chart is a well-established tool that allows the
understanding of the average convergence behavior of the mutual
information (MI) in iterative soft-information processing systems
\cite{tenBrink00}. Fig.~\ref{fig:exit} shows the results of an EXIT
chart analysis on various competing schemes. A $2\times2$
SM-MIMO-OFDM system is used for this, and an SNR of 14 dB is chosen.
$I_{A_1}$ and $I_{E_1}$ measure the MI at the input and output of
the demapper, respectively, whereas $I_{A_2}$ and $I_{E_2}$ are the
respective MI at the input and output of the decoder. At the next
iteration stage, $I_{E_1}$ becomes $I_{A_2}$ and $I_{E_2}$ turns to
$I_{A_1}$.

In the figure, the top-most curve indicates the average transfer
function of MI through the demapper and the bottom-most curve is the
same function for the decoder. Both the demapper and decoder EXIT
chart curves correspond to Gaussian-distributed input LLRs, and the
demapper EXIT curve is also based on the assumption of perfect
channel estimation. The stair-case MI plots represent actual MI
measured during IDD simulation runs and shows how the MI improves
through the iterative process for three different channel estimation
schemes. The gap between each stair-case MI trajectory and the
demapper EXIT curve represents the performance loss due to
imperfect-CSI. The solid stair-case line represents the proposed
channel estimation algorithm. The dashed-line (labeled ``Song")
corresponds to the Kalman channel estimator of \cite{Song02} applied
to the conventional-IDD setting (non-pipelined IDD with a demapper
utilizing the noise-variance update of (\ref{eq:upvar}) with channel
estimation using only the decoder output decision). The dotted line
is for the demapper utilizing only the preamble-based initial
channel estimation (following the IEEE 802.11n format, where a fixed
number of initial preamble symbols in the high-throughput long
training field is utilized). For the proposed scheme, the MI
trajectory measurement is taken from the last demapper in the
pipeline, as the last demapper block best reflects the quality of
the final decisions.
\begin{figure}
\centering
\includegraphics[width=3.5in]{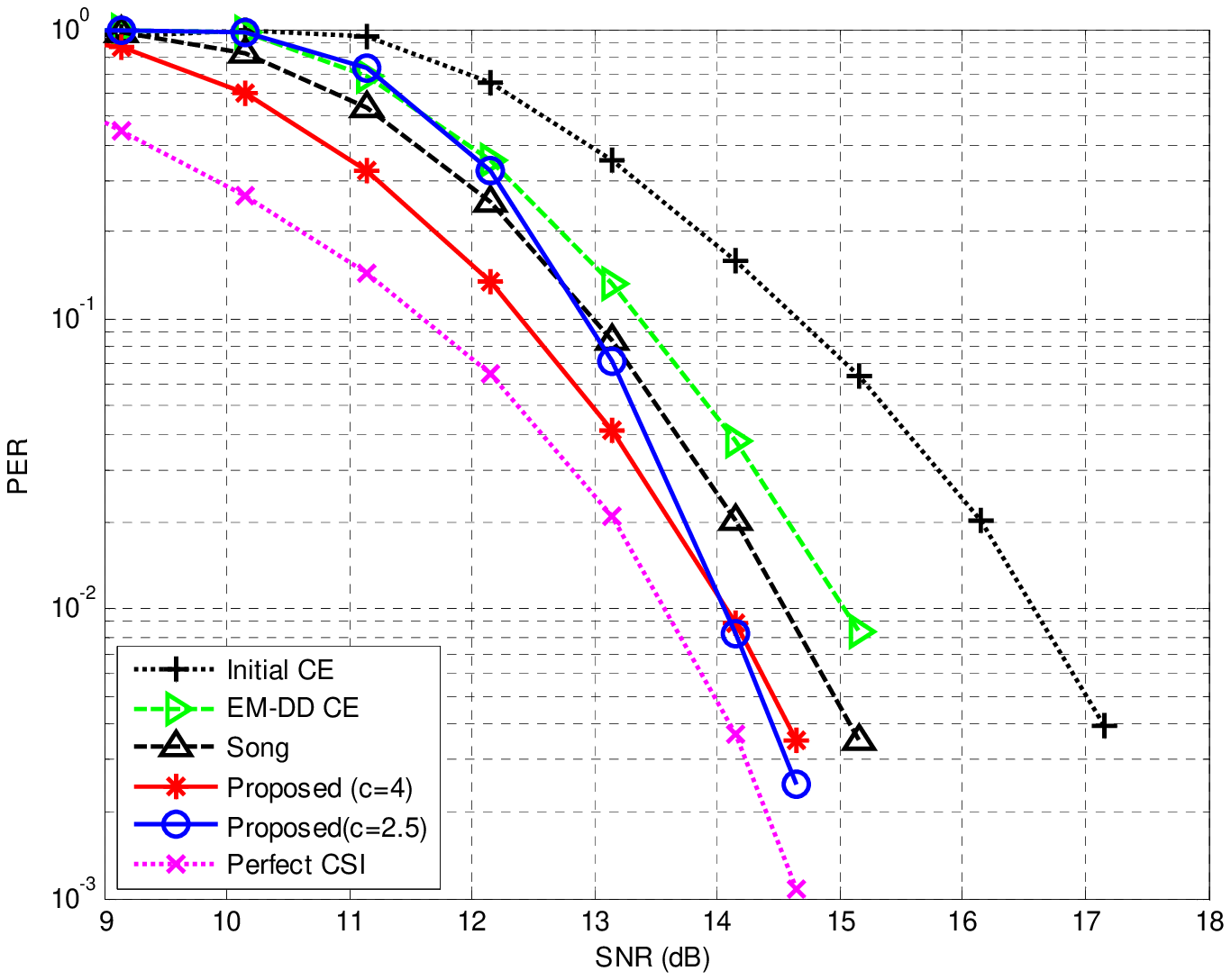}
\caption{PERs with different channel estimators: 2x2 SM-MIMO-OFDM (7
iterations)} \label{fig:MCS11_mat}
\end{figure}
It is clear that the proposed punctured-feedback Kalman estimation
with pipelined-IDD shows superior MI convergence characteristics.
The scheme of \cite{Song02} fails to improve MI beyond nine
iterations. With the demapper utilizing only initial channel
estimation, the trajectory fails to advance earlier in the
iteration.

Fig.~\ref{fig:MCS11_mat} shows PER performances of the receivers
with different channel estimators in the $2\times2$ SM-MIMO-OFDM
system. Seven iterations are applied beyond which the iteration gain
is plateaued. The performance gap between perfect CSI and
preamble-based initial CE only is nearly 3 dB at low PERs. It can be
seen that at low PER the proposed estimator almost compensates for
the loss due to imperfect-CSI when the threshold parameter is set at
$c=2.5$. Although the performance with small $c$ has inferior
performance at low SNRs, the proposed Kalman CE curve with $c=2.5$
crosses the $c=4$ curve as SNR gets higher. The large $c$ is
effective in averaging noise in low SNR, but allows relatively large
correlated errors. As expected from the EXIT chart analysis results,
the Kalman estimator of \cite{Song02} that utilizes only the decoder
output in a non-pipelined setting does not perform as well. As one
of the algorithms considered for comparison, the decision-directed
EM estimator (referred to as EM-DD here) introduced as a variant of
the EM estimator in \cite{Wautelet07} is applied with $
\hat{\bf{h}}_{o,n}^{(r)} = \left( {\widetilde{\bf{S}}_n^H
\widetilde{\bf{S}}_n } \right)^{ - 1} \widetilde{\bf{S}}_n^H
{\bf{z}}_n^{(r)}$. In addition, the EM estimate is blended with the
preamble-based channel estimate by a combining method (i.e.,
$\hat{{h}}_n^{(t,r)} = a_n\hat{h}_{preamble}^{(t,r)}  +
b_n\hat{h}_{o}^{(t,r)}[n]$) \cite{Kobayashi01}. A method to find the
combining coefficients $a_n$ and $b_n$ is discussed in
\cite{Kobayashi01}. The EM noise variance update method is presented
in \cite{Wautelet07} as $ \widehat{\mathcal N}_o [n] = {1}/{{N_r N_d
}}\sum\limits_{r = 1}^{N_r } {\sum\limits_{d = 0}^{N_d - 1} \left(
{z_n^{(r)}  - \widetilde{\bf{S}}_n \hat{\bf{h}}_n^{(r)} } \right)^*
\left( {z_n^{(r)}  - \widetilde{\bf{S}}_n \hat{\bf{h}}_n^{(r)} }
\right)}$. As can be seen, this scheme also does not perform as well
as the proposed algorithm.
\begin{figure}
\centering
\includegraphics[width=3.5in, height=2.9in]{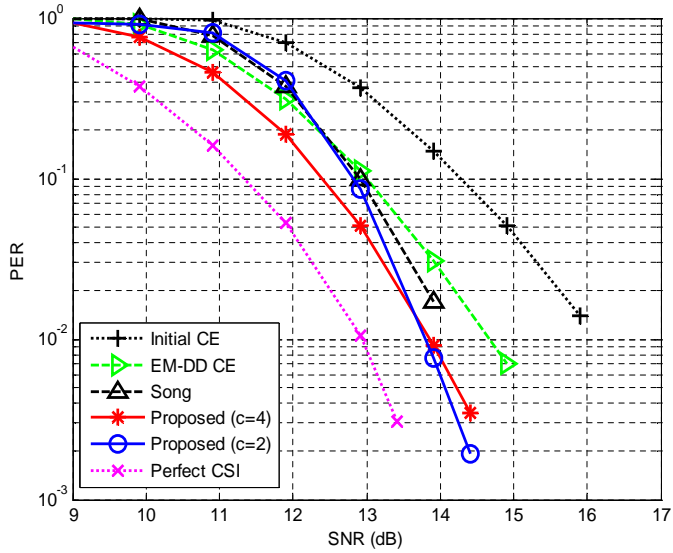}
\caption{PERs with different channel estimators: 3x3 SM-MIMO-OFDM (7
iterations)} \label{fig:MCS19_mat}
\end{figure}

\begin{figure}
\centering
\includegraphics[width=3.5in, height=2.9in]{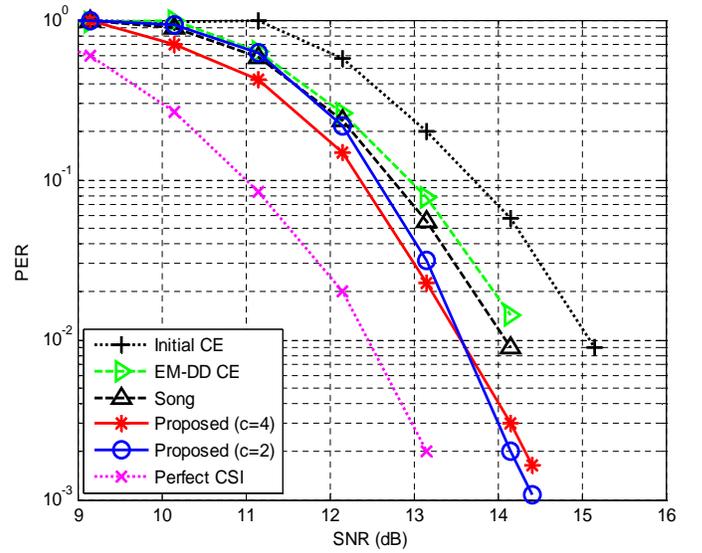}
\caption{PERs with different channel estimators: 4x4 SM-MIMO-OFDM (9
iterations)} \label{fig:MCS27_mat}
\end{figure}

Fig.~\ref{fig:MCS19_mat} and Fig.~\ref{fig:MCS27_mat} show PER
curves for $3 \times 3$ SM-OFDM-OFDM  and $4 \times 4$ SM-OFDM-OFDM
systems, respectively. These figures tell a consistent story.
Namely, the initial-CE-only scheme suffers about a 3dB SNR loss
relative to the perfect CSI case. The proposed schemes close this
gap significantly, outperforming both the Kalman-based algorithm of
\cite{Song02} and the EM-based algorithm of \cite{Wautelet07}. As
for the proposed channel estimation scheme, a more aggressive
puncturing (corresponding to a lower $c$ value) tends to give a
lower PER as SNR increases.
Before finishing this section, we briefly mention complexity. For
all considered channel estimation schemes - the proposed, the Song
method and the EM-DD scheme - implementation complexity largely
arises from the matrix inversion operation. All schemes require
matrix inversions of the same dimension. Consequently, the proposed
method and the Song method require complexity that roughly grows as
$2N_r \times O(N_t^3)$ whereas the EM-DD requires complexity
proportional to just $O(N_t^3)$. This is due to the consequence that
both our method and the Song method require matrix inversion for
each receive antenna, whereas the EM-DD method needs matrix
inversion just once and can be used for all receive antennas. The
factor 2 accounts for the fact that two matrix inversions are
required for each update of the Kalman gain in the proposed and Song
methods.

\section{Conclusions}\label{con}
A sequential soft-decision-directed channel estimation algorithm for
MIMO-OFDM systems has been proposed for the specific pipelined
turbo-receiver architecture. The algorithm deals with observation
sample sets with varying levels of reliability. In coping with
decision errors that propagate in the pipeline, we have introduced a
novel method of innovating a correlated observation sequence via
puncturing. Based on the refined innovation sequence, a Kalman-based
estimator has been constructed. The proposed algorithm establishes
improved Kalman-based channel estimation where the traditional
innovations approach cannot create a true innovation sequence due to
soft-decision error propagation. The EXIT chart, MSE analysis and
PER simulation results have been used to validate the performance
advantage of the proposed channel estimator.

\begin{IEEEbiography}[{\includegraphics[width=1in,height=1.25in,clip,keepaspectratio]{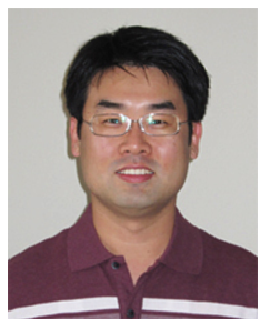}}]
{Daejung Yoon} received the B.S. and M.S. degrees in electrical
engineering from the Kyungpook National University, Daegu, Korea, in
2003 and 2005, respectively. Currently, he is working toward the
Ph.D. degree in electrical engineering at the University of
Minnesota at Minneapolis. Since 2009, he has worked as a senior
engineer in Samsung Information Systems America, San Jose, CA. His
research interests are in the general areas of communication
systems, advanced signal processing for digital communications and
hard-disk signal processing.
\end{IEEEbiography}

\begin{IEEEbiography}[{\includegraphics[width=1in,height=1.25in,clip,keepaspectratio]{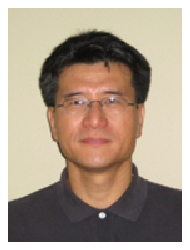}}]
{Jaekyun Moon} received a BSEE degree with high honor from the State
University of New York at Stony Brook in 1984 and M.S. and Ph.D.
degrees in the Electrical and Computer Engineering Department at
Carnegie Mellon University in 1987 and 1990, respectively. In 1990,
he joined the faculty of the Department of Electrical and Computer
Engineering at the University of Minnesota, Twin Cities, as an
Assistant Professor. He was promoted to a Tenured Associate
Professor in 1995 and then to a Full Professor in 1999. Prof. Moon's
research interests are in the area of channel characterization,
signal processing and coding for data storage and digital
communication. His recent interests are in coding and equalization
for interference-dominant channels. Prof. Moon was selected to
receive the IEEE-Engineering Foundation Research Initiation Award in
1991 and received the NSF Reseacrh Initialtion Award in the same
year. He received the 1994-1996 McKnight Land-Grant Professorship
from the University of Minnesota. He also received the IBM Faculty
Development Awards as well as the IBM Partnership Awards. He was
awarded the National Storage Industry Consortium (NSIC) Technical
Achievement Award for the invention of the maximum transition run
(MTR) code, a widely-used error-control/modulation code in
commercial storage systems. Prof. Moon served as Program Chair for
the 1997 IEEE Magnetic Recording Conference. In 1998, he was a
Visiting Professor at Seoul National University. He is also a past
elected Chair of the Signal Processing for Storage Technical
Committee of the IEEE Communications Society. In 2001, he co-founded
Bermai, Inc., a fabless semiconductor start-up, and served as
founding President and CTO. From 2004 to 2007, Prof. Moon worked as
a consulting Chief Scientist for DSP Group, Inc. He served as a
guest Editor for the 2001 IEEE J-SAC issue on Signal Processing for
High Density Storage. He also served as an Editor for IEEE
Transactions on Magnetics in the area of signal processing and
coding for 2001-2006. While on leave from the University of
Minnesota, he worked as Chief Technology Officer at Link-A-Media
Devices Corp. in 2008. Prof. Moon joined KAIST as a Full Professor
in 2009. He is an IEEE Fellow.
\end{IEEEbiography}

\end{document}